\documentclass[
superscriptaddress,
amsmath,
amssymb,
preprint,
]{revtex4-1}

\usepackage{lineno}
\usepackage{graphicx}
\usepackage{dcolumn}
\usepackage{bm}
\usepackage[utf8]{inputenc}
\usepackage[T1]{fontenc}
\usepackage{mathptmx}
\usepackage{etoolbox}
\usepackage[version=4,arrows=pgf-filled,
mathfontname=mathsf]{mhchem}
\makeatletter
\def\@email#1#2{%
 \endgroup
 \patchcmd{\titleblock@produce}
  {\frontmatter@RRAPformat}
  {\frontmatter@RRAPformat{\produce@RRAP{*#1\href{mailto:#2}{#2}}}\frontmatter@RRAPformat}
  {}{}
}%
\makeatother
\begin{document}

\preprint{}

\title{On Simulating Thin-Film Processes at the Atomic Scale Using Machine Learned Force Fields}
\author{S. Kondati Natarajan}
\affiliation{ 
Synopsys Denmark ApS, Fruebjergvej 3, 2100 Copenhagen, Denmark
}%
\email{sureshko@synopsys.com.}
\author{J. Schneider}%
\affiliation{ 
Synopsys Denmark ApS, Fruebjergvej 3, 2100 Copenhagen, Denmark
}%
\author{N. Pandey}%
\affiliation{Synopsys India Pvt. Ltd., DLF Tech Park, Noida, UP - 201307, India}
\author{J. Wellendorff}%
\affiliation{ 
Synopsys Denmark ApS, Fruebjergvej 3, 2100 Copenhagen, Denmark
}%
\author{S. Smidstrup}%
\affiliation{ 
Synopsys Denmark ApS, Fruebjergvej 3, 2100 Copenhagen, Denmark
}%

\date{\today}
\begin{abstract}
Atomistic modeling of thin-film processes provides an avenue not only for discovering key chemical mechanisms of the processes but also to extract quantitative metrics on the events and reactions taking place at the gas-surface interface. Molecular dynamics (MD) is a powerful computational method to study the evolution of a process at the atomic scale, but studies of industrially relevant processes usually require suitable force fields, which are in general not available for all processes of interest. However, machine-learned force fields (MLFF) are conquering the field of computational materials and surface science. In this paper, we demonstrate how to efficiently build MLFFs suitable for process simulations and provide two examples for technologically relevant processes: precursor pulse in the atomic layer deposition of \ce{HfO2} and atomic layer etching of \ce{MoS2}.

\end{abstract}

\noindent\fbox{%
    \parbox{\textwidth}{%
        This article may be downloaded for personal use only. Any other use requires prior permission of the author and AIP Publishing. This article appeared in (S. Kondati Natarajan, J. Schneider, N. Pandey, J. Wellendorff, S. Smidstrup. On simulating thin-film processes at the atomic scale using machine-learned force fields. Journal of Vacuum Science \& Technology A 2025, 43 (3)) and may be found at (\href{https://doi.org/10.1116/6.0004288}{https://doi.org/10.1116/6.0004288})

        This article is distributed under a Creative Commons Attribution-NonCommercial-NoDerivs 4.0 International (CC BY-NC-ND) License.
    }%
}

\maketitle

\section{\label{intro} Introduction}
Continuous down scaling of critical dimensions in semiconductor devices has reached a fundamental bottleneck due to atomic size limitation.\cite{theis2017end} The answers to this challenge are new materials, 3D integration and new devices which brings along a sleuth of challenges in terms of device fabrication. One of the main challenges is: how to deposit or etch a thin layer of material (nm scale) on a high aspect ratio structure? Atomic layer processing came to the rescue and has been the key enabling technology for continuing the trend of increased transistor densities.\cite{george2010atomic,kanarik2015overview} These emerging technologies offer atomic level control over the thickness of the film being deposited or etched in one process cycle. Since the margins for error have become very small, careful investigation of the processes in terms of the yield and purity of deposited material has become imperative. Atomic scale modeling provides a fundamental understanding of the reactions and events taking place at the gas-substrate interface during material fabrication process.\cite{ovanesyan2019atomic,kunene2022review} This knowledge can be harnessed in improving existing processes or developing entirely new processes without resorting to the expensive trial and error approach.

Three levels of atomic scale modeling are possible for simulating surface processes: thermochemical models, kinetic models and molecular dynamics (MD).\cite{quantumatk2024} Thermochemical models~\cite{kreitz2025unifying} are typically used for screening key reactions in a process from a predefined list of possible reactions by comparing their Gibbs free energies. These models are quick to evaluate as we focus only on the states of the system (reactant, product) and not the pathway connecting these states. Such models have allowed prediction of theoretical process windows and process selectivity using modest computational resources.\cite{kondati2018modeling, mullins2020self} In kinetic modeling,\cite{de2016review} a database of the screened reactions from the thermochemical studies is created and their reaction pathways are explicitly calculated. From the reaction pathways, an activation energy for the reactions as well as their rates are extracted at the process conditions using transition state theory and appended to the database. Once such a database is available, then the process can be simulated at the mesoscopic level using approaches such as kinetic Monte-Carlo.\cite{shirazi2014atomistic} In MD,\cite{ciccotti2022molecular} however, no predefined reaction database or experimental data are required; instead the process is simulated by explicitly taking into account the time-evolution of the system at the gas-substrate interface under the reactor conditions. 

MD simulation allows us to track the gas species as they approach the substrate material and collide with it. Upon collision, the gas species may either stick to the surface or reflect back into gas phase. When the gas species sticks to the surface, they may remain either dormant or activate chemical reactions and form by-products. While MD can be performed using ab initio methods,\cite{weckman2015first,sangiovanni2018ab,liu2022self} they are often too expensive for long length and time scales. Classical force fields, which use empirical functional forms to describe the interactions between atoms in a system, have been used in molecular dynamics based atomic layer process simulations.\cite{athavale1995molecular,vella2022molecular} While empirical force fields are computationally efficient, they are not trivial to generate. This is because the functional forms of the empirical potential energy surfaces have to be chosen specifically for proper representation of the interactions present in the system under study.\cite{liu2019balance} In recent times, machine learning has been used to train a new class of force fields which do not rely on empirical functional forms.\cite{mlff_compare, unke2021machine, batatia-2024, kulichenko2024data, wang2024machine} These are mathematical models fitted to ab initio data that provide the best of both worlds: accuracy at the ab initio level and performance at the classical force field level. 

Hitherto, different flavors of MLFFs have been published in the literature. Two popular classes of MLFFs are available which use either linear regression (LR)\cite{mtp_intro, bartok2010gaussian, thompson2015spectral} or neural networks (NN)\cite{behler2007generalized, scarselli2008graph, zhang2018deep, schutt2018schnet, chmiela2018towards, batatia2022mace} for interpolation. 
We use moment tensor potential (MTP), a LR model, for training MLFFs in this work as they have been identified as the most efficient method for a given accuracy of predicting the reference method.\cite{mlff_compare}
MLFFs have been applied to study different atomistic systems including molecules,\cite{natarajan2015representing, chmiela2023accurate} bulk systems,\cite{owen2024complexity} solid-liquid interfaces,\cite{natarajan2016neural,Liu2024onthefly} 2D materials,\cite{marmolejo-tejada-2022} metal organic frameworks\cite{wieser-2024} and so on. However, only a handful of such studies were dedicated towards material growth simulations like deposition of amorphous carbon\cite{caro2020amorphous,zhang2024active}, cdTe\cite{li2024elucidating}, Aluminium\cite{chapman2020nanoscale} and carbon nanotubes\cite{hedman2024dynamics}, to name a few. In the above works, simulations are performed by depositing atoms at the growth surface directly without explicitly considering the chemistry of precursor molecules. There has also been some recent published works where MLFFs were used in continuous deposition~\cite{li2022towards} and continuous etching simulations~\cite{hong2024atomistic} that included chemical effects of the reactant species interacting with the substrate material. However, to the best of our knowledge, the authors are not aware of any MLFFs applied to investigate atomic layer deposition or atomic layer etch processes involving surface reactions of gas molecules or plasma species. In this paper, we discuss a promising methodology to generate a state-of-the-art MLFF of the type MTP and apply it to study dynamic simulations of reactions taking place during: precursor pulse in the atomic layer deposition of \ce{HfO2} and atomic layer etching of \ce{MoS2}.

\section{\label{comp}MLFF Training Concepts}

\subsection{\label{mlff}Moment Tensor Potential}
Moment tensor potential is a class of MLFF that offers a good balance between performance and computational cost.\cite{mlff_compare} We use the version of MTP as implemented in the Synopsys QuantumATK atomic scale modeling platform.\cite{quantumatk2024,smidstrup2020quantumatk} At its core, the MTP predicts total energy of the system as a sum of atomic contributions,
\begin{equation}
E_{total}= \sum_i^N E_i.
\end{equation}
Here, $N$ is the number of atoms in the system. ${E_i}$ represents energy of an atomic environment centered at atom $i$, which can in turn be predicted as a linear combination of scalar basis functions representing the atomic environment,
\begin{equation}
E_{i}= \sum_j^M \xi_j B_j.
\end{equation}
Here, $M$ is the number of basis functions and $\xi$ are the linear (regression) coefficients. The linear (regression) coefficients can optionally be made element-specific, which results in improved accuracy for multi-element systems. In that case, there will be one set of linear (regression) coefficients for each element in the system. The scalar basis functions, $B$, are constructed from contractions of moment tensors of the form,
\begin{equation}
    M_{\mu,\nu} (n_i) = \sum_j f_{\mu}(|r_{ij}|, z_i,z_j) \textbf{r}_{ij} \otimes ... \otimes \textbf{r}_{ij}.
    \label{moment_tensor}
\end{equation}
Here, $f$ is a function describing the radial distribution of the neighboring atoms $j$ that are located within the atomic environment of atom $i$. A set of non-linear coefficients determining the shape of the radial functions are chosen which are always element-pair-dependent. The radial function $f$ also includes a cutoff function which takes in a user defined cutoff radius so that only neighbors $j$ with a cutoff radius of $r_{cut}$ from atom $i$ are used in the descriptor calculation. $\textbf{r}_{ij}$ is a vector whose magnitude is the radial distance between atoms $i$ and $j$. $\textbf{r}_{ij} \otimes ... \otimes \textbf{r}_{ij}$ is a $\nu$ dimensional tensor made out of the $\textbf{r}_{ij}$ vectors. In QuantumATK, the user has the option to directly provide the number of scalar basis functions, $B$, to describe the atomic environment. Detailed mathematical derivations of MTP are found elsewhere.\cite{mtp_intro, achar2022using}

\subsection{\label{mtp_flow}MTP Basic Training Flow}
The recipe for training an MTP starts with the preparation of training data representing the target system.\cite{kulichenko2024data} Training data includes 3D geometries of the system sampled at the relevant configuration space of interest and their properties such as energies, atomic forces and stresses calculated using a reference method (usually density functional theory (DFT)). A realistic system of interest may be quite large with 10s of 1000s of atoms in it. But, such a system size is not practical for DFT reference calculations. Therefore, smaller geometries with a few 100 atoms representing different regions of this realistic system can be used for training instead. This is possible because, in most MLFFs including MTP, the total energy of the system is expressed as a sum of atomic contributions based on their immediate chemical environment within a cutoff radius. Thus, a periodic system with lattice dimension of at least twice the cutoff radius is usually sufficient for the reference calculations. Some authors used ab initio MD to generate these training data.\cite{frank2024euclidean} This is acceptable for sampling a system in its equilibrium or narrow process conditions. However, if the need is to sample different phases of the system at a wide range of process conditions, ab initio MD quickly becomes inefficient in generating the necessary training data. In our approach, we use special protocols to generate reference geometries in order to avoid expensive ab initio MD simulations. These protocols will be described further below in this paper. Once the training geometries are generated, corresponding training data are calculated using the reference method. 

For training, a description of the 3D structures that is understandable by the ML model is needed.\cite{musil-2021} This is usually termed as descriptor or feature vector. A descriptor is nothing but a fingerprint of an atomic environment. To accurately describe an atomic environment, the descriptor should be sensitive to small changes in the relative atomic positions. Moreover, the descriptors must be invariant to permutations of like atoms in the system as well as invariant to translation and rotation of the system in 3D space. All of the above requirements are satisfied by the moment tensor descriptor described in Section~\ref{mlff}, which is used to encode the geometries in a multi-dimensional vector space.\cite{mtp_intro} The descriptor calculation can be tuned using the hyper parameters of the descriptor method, which in the case of MTP are the non-linear coefficients.  The next step is MTP training, where a linear regression model that relates descriptors to properties is generated, which can be used for inference in MD. 

\subsection{\label{dft}Reference Method}
We used Density Functional Theory as implemented in QuantumATK as the reference method which provided good accuracy at reasonable computational effort. We used norm-conserving pseudopotentials from PseudoDojo~\cite{van2018pseudodojo} to describe the core electrons and linear combination of atomic orbitals (LCAO) to represent the valence electronic states. It is important to pay particular attention to the type of material being simulated and choose appropriate reference method. For example, a multi-layer \ce{MoS2} system requires dispersion correction to be included, which is not present by default in all DFT methods. As a rule, the same DFT calculator settings must be used to generate all training data used to train MLFF. Mismatch in the calculator settings of the training data might lead to discontinuities in the potential energy surface learned by the ML model, which is undesirable. Since we included different types and sizes of systems in the training data set for process simulations, we had set the same K-point density in the periodic directions and only a gamma point in the non-periodic directions.

More details on the specific DFT settings are provided for the example cases in the results section further below.

\subsection{\label{gen}Training Approaches}
\begin{figure}[h]
    \centering
    \includegraphics[]{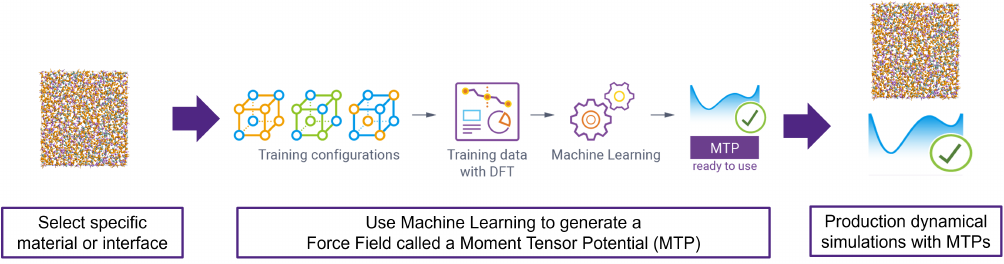}
    \caption{Schematic for batch learning.} 
    \label{fig:batch}
\end{figure}

We follow two approaches to train MTP: batch learning (BL) and active learning (AL). In batch learning, training data representing the target application system are generated first as mentioned in Section~\ref{mtp_flow}. A MTP model fitted to this training data is created and employed in production MD runs of realistic system sizes as shown in Figure~\ref{fig:batch}.

\begin{figure}[h]
    \centering
    \includegraphics[]{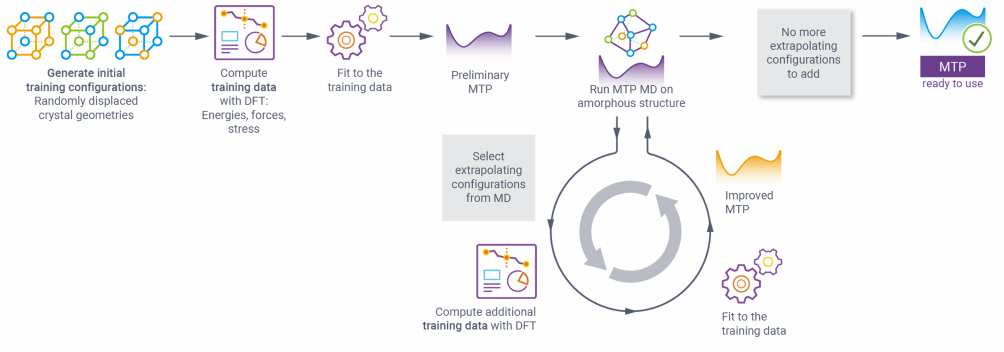}
    \caption{Schematic for active learning.}
    \label{fig:active}
\end{figure}

However, we cannot generate all relevant training data using batch learning alone. Therefore, we employ the second approach, active learning, to improve the initial training data as shown in Figure~\ref{fig:active}. Active learning starts with a preliminary reference data set, from which a preliminary MTP is generated. Then, the preliminary MTP is employed in a MD simulation of a target system at target process conditions. During the MD, a quantity determining how different the current MD snapshot is when compared to the training set is calculated every 'n' MD steps. This quantity is termed as 'extrapolation grade'.\cite{podryabinkin2017active} Note that a well trained MLFF is capable of interpolation of the potential energy surface but not extrapolation. 

So, it is important to identify when the MLFF is extrapolating as it marks the knowledge boundary for the MLFF in the configurational space of the system. To push the MLFF knowledge boundary further, we add more training data in the extrapolating regions of the configuration space. To enable this, we set two threshold values for the extrapolation grade namely, candidate threshold and retraining threshold.\cite{podryabinkin2017active} If the extrapolation grade of a MD snapshot goes above the candidate threshold, then the MD snapshot is added to a candidate list and the MD proceeds. If the extrapolation grade goes above the retraining threshold, then the MD is stopped. Thereafter, unique configurations from the candidate list are identified and their energies, forces and stresses are computed using the reference method. These new data are added to the preliminary training data to create a new improved training data set. An improved MTP is then fitted to this data and further employed in the MD. This procedure is repeated until no extrapolated configurations are identified in the MD. This procedure automatically improves the MTP by adding missing training data from the regions of configuration space that are relevant for the target application.

\subsection{\label{domains}Training Data Domains}
\begin{figure}[ht]
    \centering
    \includegraphics[]{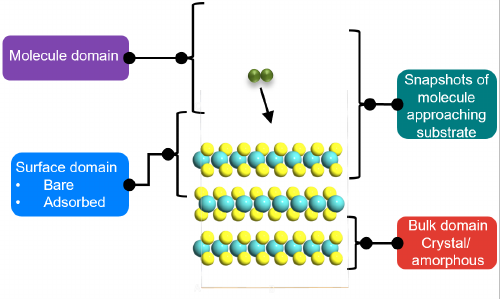}
    \caption{Schematic showing different training data domains needed to be sampled for training data generation. Four data domains are identified, namely molecule domain, bulk domain, surface domain and molecule-surface interface domain.}
    \label{fig:domains}
\end{figure}

For training an MLFF applicable to simulate processes such as ALD or ALE, different training data domains of the system are needed to be represented in the dataset as listed below (also in Figure~\ref{fig:domains}):
\begin{itemize}
    \item Molecule domain,
    \item Bulk domain,
    \item Surface domain,
    \item Molecule-surface interface domain.
\end{itemize}

\begin{figure}[ht]
    \centering
    \includegraphics[]{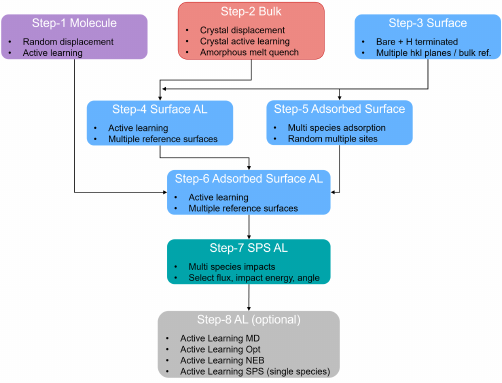}
    \caption{Data flow in the MTP training procedure for ALD and ALE process simulation. Eight step procedure is laid out in 5 levels. The steps are split into levels to enhance the efficiency of training data generation. Steps in each level can be executed in parallel as they are independent of each other. First level has three steps where the three different domains, namely molecules, bulk and surface are sampled independently. The data generated in first level is passed on to the second level. In the second level there are two steps, namely surface active learning and adsorbed surface generation, which can be run in parallel. The third, fourth and fifth levels have just one step each since they are dependent on the data from their preceding level. Step-6 in third level uses all the data generated so far in previous levels to perform active learning of adsorbed surfaces. Step-7 is used to perform active learning of gas surface collisions. Step-8 is optional and can be used to improve the training data for specific process conditions of interest.}
    \label{fig:flow}
\end{figure}

For all the domains, a preliminary dataset and final dataset are generated using automatic structure generation protocols and active learning, respectively. This data flow is described in an 8 step procedure shown in Figure~\ref{fig:flow}. This hierarchical approach allows for data sharing among MTP projects as well.

In the data flowchart, the training steps are arranged in 5 levels where the steps in each level are independent of one another, so they can be run in parallel. The first level contains three steps:  
\begin{enumerate}
	\item Step-1 Molecule: This step samples the molecule domain. Initial representative geometries of molecules involved in the process such as the precursor, reactant, plasma species, gas by-products are generated either by building them using graphical tools like the Nanolab in QuantumATK or by taking them from literature and/or online databases. Then, these initial  geometries are randomly perturbed and their corresponding reference data are added to the preliminary molecule training dataset. Active learning is then performed on the selected molecule configurations to stabilize them and also update the corresponding molecule specific training data set. The MD settings are chosen to cover the final process conditions.
	\item Step-2 Bulk: The substrate material unaffected by the surface reactions is termed 'bulk domain' and it can either be crystalline or amorphous depending on the process requirements. After selecting the relevant crystalline phase configurations of the substrate material, a set of perturbed configurations are first generated, followed by active learning. In the case of amorphous bulk system, a melt-quench MD is used in active learning with suitable annealing temperature and cooling rate. Melt-quench MD is a sequential MD procedure typically used to generate amorphous structures of a material. In this procedure, the system is first equilibrated at the room temperature using NVT ensemble. Then, the equilibrated system is slowly heated up to temperatures above its melting point using NPT ensemble. Following that, an annealing run is performed on the molten system at the same high temperature using NVT ensemble. In the next step, the molten system is cooled to room temperature using NPT ensemble at chosen cooling rate. This allows the system to be trapped in the amorphous state. Finally, the amorphous system is equilibrated at room temperature. The first two steps, however, can be excluded when one generates initial molten structure of the material using tools like PACKMOL.\cite{martinez-2009}
	\item Step-3 Surface: This step is part of surface domain sampling. From the selected crystal configurations, surfaces can be generated by cleaving them along relevant Miller planes. The surfaces can be generated either with dangling bonds or H passivation as per requirement. Active learning of surfaces is moved to next step because we also need the final bulk training data from step-2 from this level.
\end{enumerate}

In the second level, we have two steps that can be run in parallel which contribute to sampling the surface domain:
\begin{enumerate}
	\setcounter{enumi}{3}
	\item Step-4 Surface AL: The input data for this step are the combined output data from step-2 and step-3. Surface models representing the substrate material used in the ALD/ ALE process of interest are generated and used as input geometries for the active learning MD simulation. 
	\item Step-5 Adsorbed Surface: Geometries where selected molecular species are adsorbed on the optimized surface models generated in step-3 are generated to form the initial adsorbed surface training data.
\end{enumerate}

Third, fourth and fifth levels have only one step each:

\begin{enumerate}
	\setcounter{enumi}{5}
	\item Step-6 Adsorbed Surface AL: This is the final step for surface domain sampling. Output data from step-1, step-4 and step-5 are combined to create the input data for this step. The main goal of this step is to stabilize molecule adsorbed surface geometries in active learning simulations.
	\item Step-7 SPS AL: The molecule-surface domain is the most important domain for this application as it includes the geometries where the molecule is approaching the surface and adsorb on it. Surface process simulation (SPS) is the framework in QuantumATK for simulating gas species impacting a given surface as a series of MD simulation using user-defined process conditions such as substrate temperature, inlet nozzle temperature for gas species, impact energy, impact angle, flux ratio of incoming species and dosage. SPS will be explained in detail in the next section. At the end of this step, we would have generated over 90 percent of the data needed to train a good MTP for the ALD/ ALE process simulation.
	\item Step-8 AL (optional): This is an optional active learning step to fine-tune the training data towards the targets of the process simulation. This includes applying active learning in geometry optimization to identify local minima, in nudged elastic band calculations to estimate critical reaction barriers, in MD simulations to stabilize systems with needed composition of elements, and in SPS to ensure the MTP is stable at all relevant process conditions.
\end{enumerate}

At the end of this 8 step process, we will have an MTP that is ready for production runs. %

\subsection{Surface Process Simulation}
\begin{figure}[ht]
    \centering
    \includegraphics[]{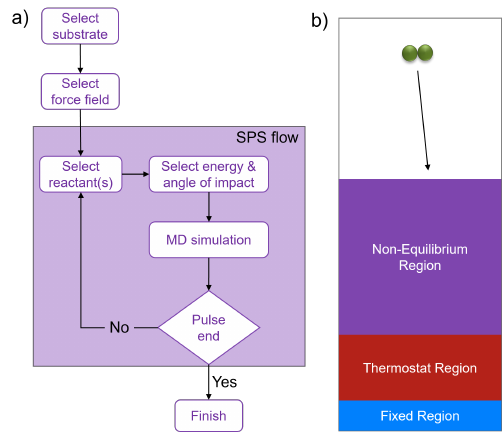}
    \caption{SPS flowchart showing the simulation procedure in panel (a). System setup for MD in SPS is shown in panel (b).}
    \label{fig:sps_flow}
\end{figure}

Surface process simulation (SPS) is a method available in QuantumATK to perform sequential MD simulations\cite{schneider2017atk} where each MD simulation corresponds to one reactant species impinging on the surface. A typical flowchart of SPS is given in Figure~\ref{fig:sps_flow}a.
After selecting the substrate material and the force field, an SPS flow can begin. The reactant species can be either one or many depending on the process under study. The choice of impact energy and impact angle is then made. For thermal processes, the internal energy of the incoming molecule is set to the nozzle temperature. So, in this case the kinetic energy will be rather low in the order for few 10s of meV and the angle of approach will be random. Note that the incoming species is initialized at a user-defined distance from the top surface of the substrate before the MD simulation begins. For plasma sources, a well defined energy and angle distribution of the participating plasma species can be provided based on plasma chamber simulation. Based on the flux of the incoming species, the number of impacts is chosen for the given surface area of the substrate. Usually a smaller surface area is chosen for simulation (few nm$^2$) when compared to the experiments, thus we can assume the molecule impacts are disconnected in time when approaching this small surface area. This justifies the use of sequential impacts of the incoming species on the substrate to simulate these processes. A special mixed ensemble setup is needed for the MD in surface process simulations. For this, the substrate is split into three regions: a fixed region, a thermostat region and a non-equilibrium region as shown in Figure~\ref{fig:sps_flow}b. Fixed region represents the bulk region of the system which is assumed to be unaffected by the impacts taking place at the surface. Thermostat region is used to regulate the temperature of the substrate. The region above the thermostat region is treated in a microcanonical way since the molecule-surface collision is a non-equilibrium process. The impact simulation is performed in three stages: impact stage, relaxation stage and equilibration stage. The incoming species has the maximum kinetic energy during the impact stage as it travels through the vacuum towards the surface. Therefore, this stage needs to be sampled finely in MD using a small time step. The simulation time at this stage depends on the kinetic energy of the incoming species.  
The other stages exist to avoid sampling the system evolution at the same finer time step after impact and it is also not needed since the velocities would have dropped after the impact due to momentum exchanges between the impacting species and the surface atoms. Relaxation stage is also an MD run with a larger time step offering approximately 5$\times$ – 10$\times$ speed up compared to the impact stage. The final stage is a long time equilibration using time-stamped force biased Monte-Carlo\cite{neyts2014combining, mees2012uniform} with a potentially 50$\times$ – 100$\times$ speed up. In the next section we will take a look at two examples where SPS is used to simulate atomic layer processes using custom trained MTPs.

\section{\label{results} Case Studies}
\subsection{\label{hfo2} Precursor pulse in \ce{HfO2} ALD Process}
\ce{HfO2} is a technologically important high-$\kappa$ dielectric material. ALD of \ce{HfO2} is possible by many different chemistries for both precursor and co-reactant pulses. One of the ALD process for \ce{HfO2} is to use \ce{HfCl4} as the precursor and use either one of \ce{H2O} or \ce{O3} or \ce{O4} or \ce{Hf(mmp)4} as co-reactant.\cite{alddatabase} Our goal for this example is to train an MTP to simulate the self-limiting reaction of the precursor (\ce{HfCl4}) pulse on the \ce{HfO2} surface explicitly at room temperature. We would like to note that for this example, we did not include any H coverage on the surface. H pre-covered surfaces will be considered in our future studies. Moreover, \ce{HfO2} ALD is a low energy process, therefore, we considered predominantly crystalline substrate material and surface adsorption of the molecule. However, this is not a restriction of the MTP model or the training methodology as we will discuss ALE of \ce{MoS2} later which is a high energy process.

\subsubsection{MTP Training and Quality}
For the reference calculator, we used generalized gradient approximation of DFT and Perdew-Burke-Ernzerhof functional~\cite{perdew1996generalized} for the exchange-correlation part of the electronic interaction. A PseudoDojo medium basis set was used to describe the valence electrons while the core electrons are described with norm-conserving pseudopotentials. We used a K-point density of 4/\r{A} in the periodic directions and only a gamma point along the non-periodic directions (also for molecules). 

\begin{table}[h]
\caption{\label{tab:hf02mtpstr} List of number of structures from different data domain used in training the MTP describing ALD of \ce{HfO2}. 
}
\begin{ruledtabular}
\begin{tabular}{lr}
Domain&Number of Structures\\
\hline
Molecule &141\\
Bulk &540\\
Surface  &90\\
Adsorbed Surfaces  &  676\\
&\\
Total & 1447\\
\end{tabular}
\end{ruledtabular}
\end{table}

For generating the MTP, we utilized 734 basis functions to describe the atomic environments within a 4 \r{A} cutoff radius. The training dataset includes 1447 structures including molecules, bulk, surface and adsorbed surface geometries as shown in Table~\ref{tab:hf02mtpstr}. Note that the geometry count includes both randomly displaced geometries and those found in AL runs. In the molecular domain, we included 141 geometries with 1, 4 and 6 \ce{HfCl4} molecules in the simulation box. In the bulk domain, we sampled 540 monoclinic \ce{HfO2} structures including 1, 4, 8, 16, 20 and 32 \ce{HfO2} units in the periodic simulation box. We added 766 surface geometries to the reference data which includes bare surfaces and molecule adsorbed surface geometries. For the extrapolation grade calculation in AL, a candidate threshold value of 1 and a retrain threshold of 3 are used. For the active learning MD simulation, we set a temperature of 300 K to sample the relevant configuration space. We reserved 10\% of the reference data for testing purposes and trained an MTP on the rest. The final MTP potential is available for use in the current release of QuantumATK.~\cite{quantumatk2024} 

\begin{figure}[ht]
    \centering
    \includegraphics[]{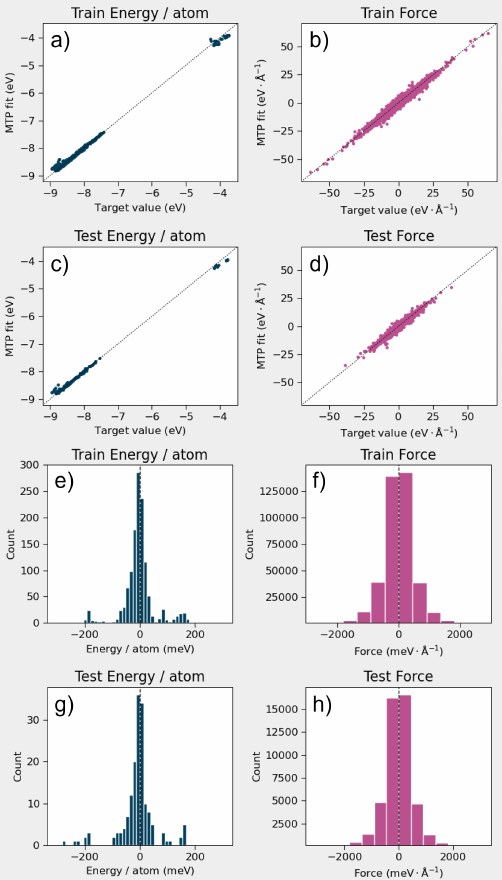}
    \caption{Quality metrics of the trained MTP for \ce{HfCl4} pulse on \ce{HfO2} ALD process. Panels a) and b) shows scatter plots comparing the training set energy and force values between MTP and target (reference DFT results). c) \& d) shows the corresponding scatter plots for the test set.  Panels e), f), g) \& h) display histograms showing the energy and force error distributions of the training and test sets.} 
    \label{fig:hfo2quality}
\end{figure}
Figure~\ref{fig:hfo2quality} and Table~\ref{tab:hfo2quality} show the quality characteristics of the trained MTP. The energy RMSE for the training set is ~60 meV/atom and force RMSE is  ~0.55 eV/Ang. It can be seen from the table that the training set and the test set give similar RMSE values for energy and forces which is also evident from the scatter plots and error histograms. The scatter plots show that the data points lie along the regression line indicating the high reproducibility of the fit. The histograms are also showing a normal distribution centred at zero as expected for both energies and forces.  

\begin{table}[ht]
\caption{\label{tab:hfo2quality} Quality metrics for the \ce{HfO2} MTP. MAE is the mean absolute error, RMSE is the root mean squared error, and R$^2$ is the coefficient of determination.}
\begin{ruledtabular}
\begin{tabular}{lccc}
Quantity&MAE [meV/atom] & RMSE [meV/atom] & R$^2$\\
\hline
Training Energy & 35.8 & 60.1 & 0.998\\
Testing Energy & 38.5 & 65.5 & 0.997\\
&MAE [meV/\r{A}] & RMSE [meV/\r{A}] & R$^2$\\
Training Forces & 378.4 & 548.3 & 0.913\\
Testing Forces & 391.2 & 574.1 & 0.910\\
\end{tabular}
\end{ruledtabular}
\end{table}

\begin{table}[htbp]
\caption{\label{tab:hf02mtpval} Comparing MTP and DFT adsorption energies of \ce{HfCl4} on different \ce{HfO2} surfaces.}
\begin{ruledtabular}
\begin{tabular}{lcr}
Surface&DFT Energy [eV] & MTP Energy [eV]\\
\hline
\ce{HfO2}(111) & -1.9 & -2.0\\
\ce{HfO2}(001) & -2.0 & -2.1\\
\end{tabular}
\end{ruledtabular}
\end{table}
As a validation study, we compared the adsorption energies of \ce{HfCl4} molecule on (111) and  (001) surfaces of \ce{HfO2} calculated with the MTP and the reference DFT method. The results are in very good agreement as shown in Table~\ref{tab:hf02mtpval}. Note that we have intentionally avoided passivating the bulk-cleaved \ce{HfO2} surface with H atoms, which would otherwise present a nucleation delay on the surface. The effects of H passivation can be included in the MTP by adding additional training data for H passivated surfaces of \ce{HfO2} in steps 3, 4, 5, 6, 7, and 8 discussed in Section~\ref{domains}. This would be addressed in our future studies. Once the training data is generated, it took about 12 minutes to train the MTP on a 16 core machine.

\subsubsection{Results}
\begin{figure}[h]
    \centering
    \includegraphics[]{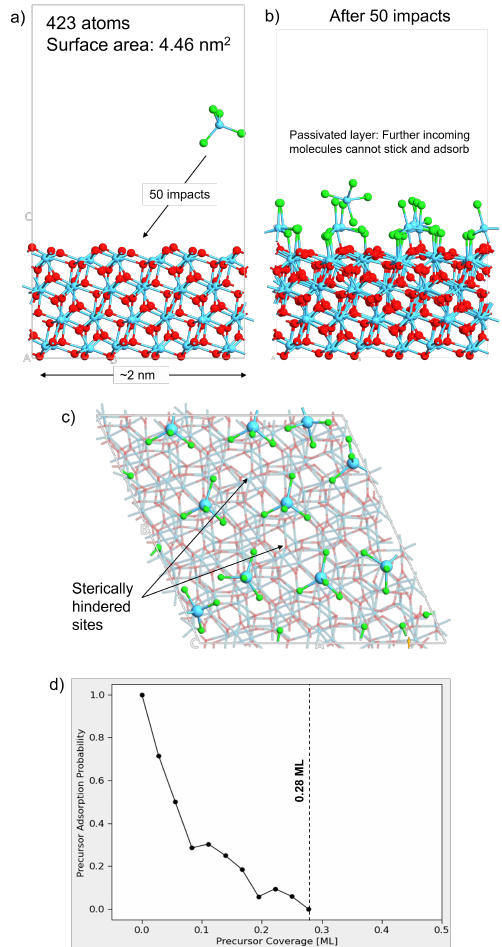}
    \caption{Results for \ce{HfCl4} impacts on \ce{HfO2}(111) surface. a) shows the reactive surface model of \ce{HfO2} used in the MD simulations, b) is a snapshot of the molecule passivated layer of \ce{HfO2} at the end of the precursor pulse, c) displays the top view of the simulation cell showing the molecular packing and sterically hindered sites, and d) shows the trend of averaged adsorption probability of \ce{HfCl4} on reactive \ce{HfO2} surface as a function of \ce{HfCl4} pre-coverage on the surface. Hf atoms are in cyan, O in red, and Cl in green. }
    \label{fig:hfo2-sps-setup}
\end{figure}

Using the final MTP, we performed a production run where \ce{HfCl4} molecules are set to impinge on the surface of \ce{HfO2}(111) with a surface area of 4.46 nm$^2$ and containing 423 atoms as shown in Figure~\ref{fig:hfo2-sps-setup}a.

We simulated 50 sequential impacts of \ce{HfCl4} molecules on the surface and repeated it 10 times with different initial conditions to account for statistical variations. For this, we performed an initial surface equilibration MD for 100 ps with a time step of 1 fs at 300 K and saved 1000 snapshots in a trajectory. Then, for each of the 10 repetitions, a surface geometry is drawn at random from the equilibrated surface trajectory and used for the 50 sequential impacts. In the MD simulations, a surface temperature of 300 K is set. The vibrations and rotations of the precursor molecule is sampled from Maxwell-Boltzmann distribution at 300 K. The translational kinetic energy is set to 0.026 eV which is the KbT value at 300 K and for the impact angle we chose a distribution centered at 0 and a standard deviation of 30${^\circ}$ with respect to surface normal direction. Each impact event is simulated in 3 stages: first MD stage for 10 ps at 0.5 fs timestep, second MD stage for 20 ps at 1 fs timestep and finally a fbMC stage for 100 ps. Figure~\ref{fig:hfo2-sps-setup}b shows a snapshot of the \ce{HfCl4} passivated surface of reactive \ce{HfO2}. It can be seen from the top view in Figure~\ref{fig:hfo2-sps-setup}c that the molecules pack the surface in an optimal way and inhibit further molecules from adsorbing on it due to sterical hindrance effects. 

Adsorption probability is defined as the ratio of the number of molecules that remain bonded to the substrate to the number of molecules impacting the surface. An adsorption probability of 1 indicates that the molecule always bonds to the surface after impact. Adsorption probability of the molecule is calculated as a function of the pre-coverage of precursor molecules on the surface ($S(\theta)$ - where $\theta$ is the surface coverage) as shown in Figure~\ref{fig:hfo2-sps-setup}d. It can be observed that the molecule sticks always when the surface is not covered, i.e. $S(\theta=0)$ = 1. The value of $S$ decreases steadily with increasing $\theta$ value and becomes 0 at $\theta=0.28$ ML. 
\newpage

The dip seen in some coverage values is due to insufficient statistics at those coverages, which could be rectified by increasing the number of repetitions. 1 ML equals to 36 \ce{HfO2} units in this simulation cell and 0.28 ML amounts to 10 Hf atoms deposited, which can be considered as a maximum theoretical estimate for this process. In literature, a stable growth rate of 1.2 - 1.5 Hf/nm$^2$ per cycle is observed~\cite{nyns2008hfo2} which is about  5.4 - 6.7 Hf atoms deposited on a similar surface area considered in our study. The difference here is probably due to the presence of surface H which could potentially limit the maximum \ce{HfCl4} adsorption. In this example case study, we have provided a recipe that can be followed to study precursor-surface interaction dynamically at the process conditions and determine theoretical maximum yield of the process. In the next section, we will discuss how a similar simulation approach can be used to investigate ALE of \ce{MoS2}.

\subsection{\label{mos2} ALE of \ce{MoS2}}
\begin{figure}[ht]
    \centering
    \includegraphics[]{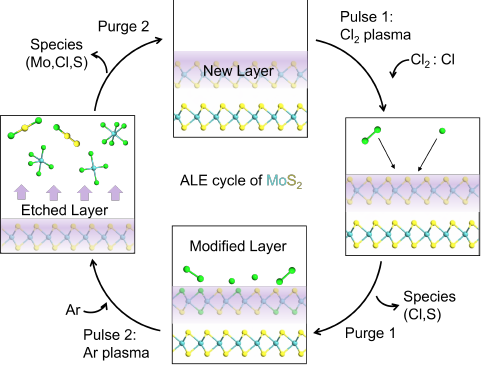}
    \caption{A schematic showing the cyclic ALE process of \ce{MoS2} using \ce{Cl2} plasma and Ar ion bombardment. Before the first ALE pulse, a new layer of \ce{MoS2} is exposed. In pulse 1, Cl plasma species enter the reaction chamber and react with the exposed \ce{MoS2} layer. In purge 1, excess chemicals are removed from the chamber leaving the surface with chemisorbed Cl species. In pulse 2, high energy Ar ions are introduced in the chamber which impact the modified surface and result in the formation of by-products. In purge 2, the by-products are removed which results in the removal of \ce{MoS2} layer.}
    \label{fig:mos2_ale_scheme}
\end{figure}

\ce{MoS2} is a 2D transition metal dichalcogenide material that is being considered as a replacement for Si as the channel material in advanced nodes. \ce{MoS2} has a bi-layered unit cell and each layer has three atomic planes: 2 S planes sandwiching a Mo plane. ALE process for \ce{MoS2} must ensure that the three atomic planes of a layer can be removed in one process cycle and this requirement makes the process more challenging. Recently, ALE of \ce{MoS2} has been demonstrated by many research groups using different chemistries \cite{aledatabase}. For this investigation we refer to the isotropic ALE process of \ce{MoS2} using \ce{Cl2} and Ar ion bombardment~\cite{kim2017atomic} as shown in Figure~\ref{fig:mos2_ale_scheme}. 

In this process, \ce{Cl2} pulse uses an inductively coupled plasma source that chemically modifies the \ce{MoS2} surface and the subsequent Ar plasma pulse removes the modified layer. It was demonstrated experimentally that 1 monolayer of \ce{MoS2} can be removed in 1 process cycle. Our goal with this case study is to understand the mechanisms taking place in the individual pulses and verify the process conditions needed for self-limiting reactions. Unlike the \ce{HfO2} case, we have trained this MTP potential to simulate both ALE pulses.

\subsubsection{MTP Training and Quality}
For the reference calculator, we chose the PBE density functional similar to the \ce{HfO2} case. We used PseudoDojo high basis set to describe valence electrons. We also increased the K-point density to 7/\r{A} along the periodic directions for the bulk and surface geometries. One thing to remember for multilayer 2D materials is that the layers are held together by non-bonding vdW interactions. Usually, DFT calculations do not include dispersion corrections out of the box, therefore these corrections must be included either with the DFT or in an ad-hoc manner while doing the MTP inference. Since the MTP is usually relatively short-ranged, we do not include the vdW corrections directly in the MTP training, but add them on top of the trained MTP model. This way, we can include the vdW interactions over longer interatomic distances via Grimme's dispersion D3 method.~\cite{grimme2011effect} We only included 2 body interaction terms for the D3 potential with a maximum neighbor cutoff of 12 \r{A}. For the MTP training, we used 750 basis functions to describe the atomic environments with a cutoff radius of 6 \r{A}.  

\begin{table}[h]
\caption{\label{tab:mos2mtpstr} List of number of structures in the training data for the MTP describing ALE of \ce{MoS2}. 
}
\begin{ruledtabular}
\begin{tabular}{lr}
Domain&Number of Structures\\
\hline
Molecule &221\\
Bulk &109\\
Surface  &282\\
Adsorbed Surfaces    &342\\
&\\
Total & 954\\
\end{tabular}
\end{ruledtabular}
\end{table}

In total, we generated 954 training configurations for this MTP and the number of geometries per training data domain is given in Table~\ref{tab:mos2mtpstr}. In this ALE process, we expected that the molecules will decompose and form gas phase by-products as the impact energies are very high when compared to the \ce{HfO2} case. Therefore, molecule geometries of the by-products (such as \ce{S2}, \ce{S2Cl2}, \ce{S2Cl}, \ce{SCl2}, \ce{SCl}, \ce{MoCl4}, \ce{MoCl5} and \ce{MoCl6}) are also included in the reference data set in addition to \ce{Cl2}. In total, we added 221 configurations in the molecule domain. We added 109 configurations in the bulk domain by sampling 2, 8 and 10 \ce{MoS2} units in a periodic box. For the surface and gas-surface interface domain, we included 624 structures in total. For the SPS AL simulations, we considered impact kinetic energies of up to 40 eV. Therefore, in the active learning MD simulations we chose a temperature in the range of 400 K to 800 K to account for the considerable local heating immediately upon impact. For this MTP potential generation, we have used all reference data for training and used active learning for validations. This way, we can reduce the reference data requirements and also reliably describe the configurational space of interest. In any case, the MTP is not expected to describe the system beyond the process conditions sampled in active learning. Once the training data is generated, it took 6 minutes to train the MTP on a 24-core machine. For testing the potential, we calculated a few properties such as lattice constant of \ce{MoS2} bilayer unitcell, cohesive energy and surface energy of \ce{MoS2}(001) using MTP and compared to the DFT results as shown in Table~\ref{tab:mos2mtpval} and we find a good agreement. The final MTP potential is available for use in the current release of QuantumATK.~\cite{quantumatk2024} 

\begin{table}[hb]
\caption{\label{tab:mos2mtpval} Comparing MTP and DFT results for lattice constant, cohesive energy and surface energy of \ce{MoS2}.}
\begin{ruledtabular}
\begin{tabular}{lcc}
Property&DFT & MTP \\
\hline
Lattice constant [\r{A}] & a = 3.18 & a = 3.19\\
 & b = 5.50 & b = 5.52\\
 & c = 12.82 & c = 13.13\\
Cohesive energy [eV] & -5.41 & -5.40\\
\ce{MoS2}(001) surface Energy [eV/\r{A}$^2$] & 0.009 & 0.007\\
\end{tabular}
\end{ruledtabular}
\end{table}

\begin{figure}[ht]
    \centering
    \includegraphics[]{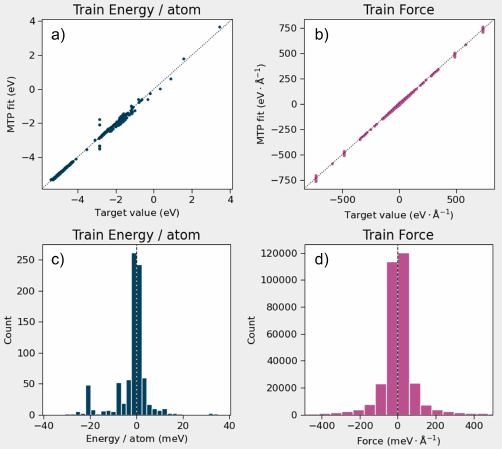}
    \caption{Quality metrics of the trained MTP for \ce{MoS2} ALE. a) \& b) Scatter plot comparing energy and force values between MTP and target (reference DFT results). c) \& d) Histogram showing the energy and force error distribution.}
    \label{fig:mos2quality}
\end{figure}

Table~\ref{tab:mos2quality} lists the energy and force errors of the final MTP compared to the reference method. We find an energy RMSE of 85.7 meV/atom and a force error of 220.3 meV/\r{A}. These values are acceptable considering the high energy/ forces in the structures included in the training data. Figure~\ref{fig:mos2quality} a and b shows the scatter plot comparing the cohesive energies and forces computed with MTP and reference DFT. Majority of the data points lie along the regression line with a R$^2$ value close to 1 indicating high reproducibility of the MTP. 
\begin{table}[ht]
\caption{\label{tab:mos2quality} Quality metrics for the MTP for ALE of \ce{MoS2}. MAE is the mean absolute error, RMSE is the root mean squared error, and R$^2$ is the coefficient of determination.}
\begin{ruledtabular}
\begin{tabular}{lccc}
Quantity&MAE [meV/atom] & RMSE [meV/atom] & R$^2$\\
\hline
Energy & 26.5 & 85.7 & 0.996\\
&MAE [meV/\r{A}] & RMSE [meV/\r{A}] & R$^2$\\
Forces & 68.9 & 220.3 & 0.999\\
\end{tabular}
\end{ruledtabular}
\end{table}

The histograms in Figure~\ref{fig:mos2quality} c and d shows that the errors are normally distributed centered at 0, which is the typical error distribution expected with MLFFs. Note that we have included very high forces in the training set as they are needed to properly include the electron repulsion due to the collision of the incoming species with the substrate atoms at high energies. Some authors use a separate empirical potential to describe the repulsive part of the PES,\cite{hong2024atomistic} however, we wanted to demonstrate with this example that MTP is capable of describing the repulsive part of the PES as well when appropriate reference structures are included in the training set. 

\subsubsection{Results}
Firstly, we will look at the results of the first ALE pulse which is the exposure of \ce{Cl2} plasma on \ce{MoS2} surface. \ce{Cl2} plasma includes several species such as ions and neutral species such as radicals, molecules and atoms.  Cl ions are species with an electron removed or added to the outer most shell of Cl atom. Then we have radicals, which are neutral species with unpaired electrons in their outer shell which can be either atomic Cl radical or molecular \ce{Cl2} radical. Ions typically have the highest kinetic energy in the plasma compared to the radical species. Radicals are more reactive than the neutral atoms or neutral molecules due to the presence of unpaired electron in their outer orbitals. 

\begin{figure}[h]
    \centering
    \includegraphics[]{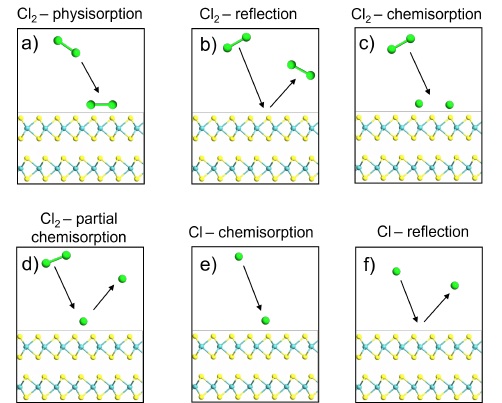}
    \caption{Possible outcomes of impinging \ce{Cl2} and Cl species. \ce{Cl2} species: physisorption (a), reflection (b), chemisorption (c), and partial chemisorption (d). Cl species: chemisorption (e), and reflection (f). }
    \label{fig:impact outcomes}
\end{figure}
In our classical MD simulations, we do not explicitly consider the electronic structure of these plasma species, however, we can differentiate them by means of their kinetic energy. We assume the ionic species would immediately be neutralized upon interaction with the surface and radical species as being simply faster than their gas phase counterparts. For comparison, thermal gas species are the slowest with a kinetic energy of about 25 meV at 300 K nozzle temperature. The kinetic energies of radicals and ions, however, depend on the plasma conditions. In the experimental reference,\cite{kim2017atomic} the authors generated the \ce{Cl2} plasma at 10 mTorr with 63 sccm of \ce{Cl2} and 18 W of a 13.56 MHz radio frequency ICP-source power. We found a peak energy of 250 meV for the Cl radicals. This is 10 $\times$ larger than that of the gas phase \ce{Cl2} species. Since an ion filter is used in the experiment, we didn't calculate the peak energies of Cl ions, but used a reference value of 15 eV for comparison. Due to the high energy, the ions are highly directional whereas the radicals and thermal gas species are more random in their angle of approach to the surface.

As a first step, we wanted to find out why a plasma source is needed in the ALE of \ce{MoS2}. For comparison, it is possible to perform ALE of Si using \ce{Cl2} gas at room temperature. To answer the above, we performed SPS simulations of 500 isolated impacts of the plasma species (Cl radical and \ce{Cl2} radical species at impact energy of 0.25 eV) and room temperature \ce{Cl2} gas (impact energy of 0.025 eV) on 4 layered pristine surface with 10 units of \ce{MoS2} in each layer (surface area of 1.4 nm$^2$) and calculated probabilities of different outcomes such as adsorption or reflection. We did not include Cl ion for this part of the study. There are 4 possible outcomes for \ce{Cl2} species impacts namely, physisorption where the molecule is adsorbed in a intact fashion, total reflection, chemisorption where the molecule adsorbs in a dissociative way as shown in Figure~\ref{fig:impact outcomes}. Two type of chemisorption are possible where either both Cl atoms are adsorbed on the surface or only one adsorbs and the other reflects back. On the other hand, for the Cl species there are only two possibilities - either adsorption or reflection.

\begin{figure}[ht]
    \centering
    \includegraphics[]{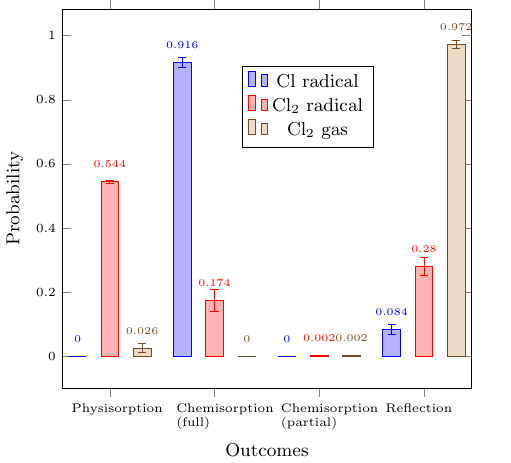}
    \caption{A bar chart comparing the probabilities of different events from Figure~\ref{fig:impact outcomes} along with error bars. The error bars are calculated by sub-sampling approach.}
    \label{fig:impact probabilities}
\end{figure}

Before the SPS run, the surface is equilibrated at room temperature before the impacts. The gas and radical species are initialized at 10 \r{A} above the top surface layer and set to approach the surface at a randomly selected angle in the range of 0$^{\circ}$ to 40$^{\circ}$ with respect to the surface normal direction. Firstly, for the thermal impacts of \ce{Cl2}, we set 5 ps for the impact stage at 0.5 fs time step followed by 5 ps of relaxation time at 1 fs time step and finally 20 ps of fbMC equilibration. For the radical species impacts, we decreased the impact durations to 2 ps as larger impact time was not needed in these cases due to fast moving species. The results are compared in a bar chart showed in Figure~\ref{fig:impact probabilities}.

It can be seen from the bar chart that the \ce{Cl2} gas at room temperature reflects over 97 \% of the time which indicates that the adsorption barrier cannot be breached comfortably by this species. Moreover, no fully chemisorbed outcomes were recorded although there was only 1 event where a partial chemisorption is seen. In contrast to this, a majority of the \ce{Cl2} radical species physisorbs (55\%) on the \ce{MoS2} surface, followed by reflection (28\%) and then chemisorption (17\%). This difference is due to the larger kinetic energy possessed by the \ce{Cl2} radical species which aids in breaching the adsorption barrier. The Cl radical on the other hand predominantly chemisorbs on the surface (92\%) and only 8\% of them reflects back. This study reveals the fact that a thermal pulse of \ce{Cl2} will not result in a successful half-cycle and that a plasma source is in fact needed.

\begin{figure}[ht]
    \centering
    \includegraphics[]{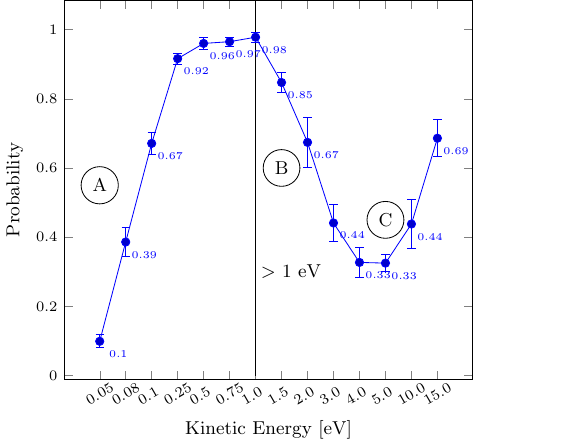}
    \caption{Plot showing the adsorption probability of incoming Cl species on \ce{MoS2} surface as a function of the kinetic energy of impact. Three regions are identified: A, B and C. A: adsorption improves with initially increasing kinetic energy of Cl species (overcome adsorption barrier), B: More reflection due to increased kinetic energy (surface phonons cannot absorb all of this energy), and C: Reflection decreases due to Cl-S exchanges}
    \label{fig:sticking probabilities}
\end{figure}

As a next study, we wanted to simulate what happens when an ion filter is not used in the experiments. For this we employed SPS to understand the adsorption probability of atomic Cl species on a similar 4-layered \ce{MoS2} surface at different impact energies in the range of 0.05 eV to 15 eV. For this study, we sampled a wider angle of impact for the species energy up to 1 eV. For higher energies, we used directional surface normal impacts. Up to 4 eV of impact energy, we simulated 3 ps of impact stage at 0.5 fs time step, followed by 3 ps of relaxation at 1 fs time step and 20 ps of fbMC equilibration. For higher impact energies, we decreased the impact stage to 2 ps and increased the relaxation time to 5 ps. Otherwise, all other settings are kept the same. The results are plotted in Figure~\ref{fig:sticking probabilities}.

\begin{figure}[ht]
    \centering
    \includegraphics[]{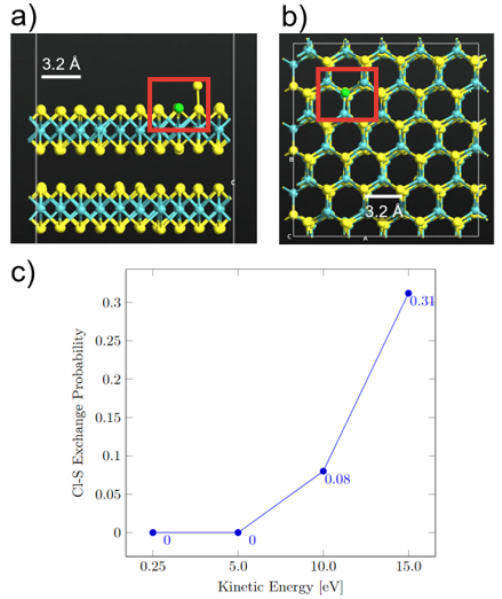}
    \caption{Panels a) and b) shows side view and top view of the surface where a Cl-S exchange mechanism is observed. Panel c) shows the probability of this exchange event at different kinetic energies of Cl species.}
    \label{fig:exchange probabilities}
\end{figure}

At a low impact kinetic energy of 0.05 eV, the Cl species did not adsorb favorably on the \ce{MoS2} surface and gets reflected back most of the time. As the kinetic energy is increased to 0.08 eV and to 0.1 eV, the probability of adsorption increases steadily as seen in region A in Figure~\ref{fig:sticking probabilities}. This is probably due to breaching the adsorption barrier at such energies. The adsorption probability starts to plateau at 92\% when reaching a kinetic energy of 0.25 eV. Note that this is also the kinetic energy of the radical species from the experimental setup. The peak adsorption probability is observed at 1 eV (98\%). Beyond 1 eV though, the adsorption probability drops steadily until 4 eV to a value of 33\% as seen in region B. This is probably due to more reflections as the high energy of the projectile species can not be absorbed effectively by the surface phonons. The adsorption probability didn't drop further but stayed at the same value of 33\% when increasing the kinetic energy to 5 eV. However, on further increasing the impact kinetic energy, the adsorption probability started increasing again suggesting that the surface is accommodating the high energy projectiles as seen in region C. On analyzing the trajectories, it is revealed that the Cl species at energies > 5 eV start damaging the surface by kicking away surface S atoms and take their place on the surface, we call this the 'Cl-S exchange' mechanism as shown in Figure~\ref{fig:exchange probabilities} a and b. This is consistent with our calculation of cohesive energy of \ce{MoS2} at 5.4 eV as shown in Table~\ref{tab:mos2mtpval}. The adsorption probability beyond 5 eV of impact energy therefore includes both surface adsorption and exchange adsorption. We also calculated the Cl-S exchange probability separately as shown in Figure~\ref{fig:exchange probabilities}c. 
It can be confirmed from the plot that these exchange reactions only take place above 5 eV and their contribution is still relatively small compared to on-surface adsorption at 10 eV, but becomes almost equally probable at 15 eV. We didn't simulate beyond 15 eV, but we expect that the ion will damage the surface further and reach bottom layers, which is undesirable. As mentioned earlier, an ion filter is used in the experiment~\cite{kim2017atomic} to only allow low energy plasma species to impact the surface so that no S atoms are removed from the surface in the Cl pulse. However, our investigation proposes that high energy ions in controlled amounts may be used to physically modify the top surface layer of \ce{MoS2} using the Cl-S exchange mechanism in a self-limiting manner. In our simulations, we did not record any sputtering events or chemical etching of the substrate species.

\begin{figure}[h]
    \centering
    \includegraphics[]{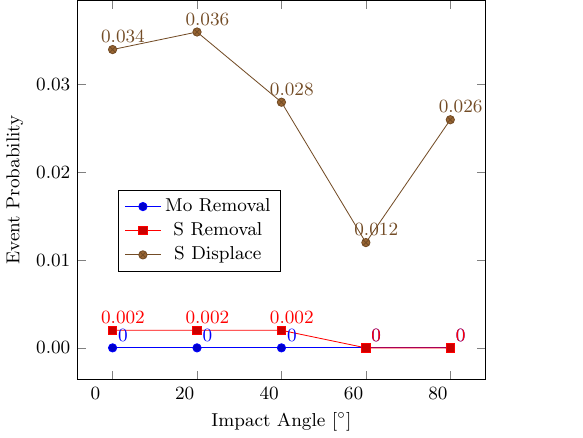}
    \caption{Plot showing the probabilities of sputtering of Mo and S along with the probability of displacing S atoms from the surface when exposed to Ar ions at a kinetic energy of 20 eV. No Mo is sputtered in the simulations at any angle of incidence.}
    \label{fig:ar_probabilities}
\end{figure}

In the next step, we wanted to verify the self-limiting nature of the Ar pulse on pristine \ce{MoS2} surface. In the experiments,~\cite{kim2017atomic} Ar ions with a peak energy at 20 eV are used, so we use the same value for the impact energy in our simulations. We used a similar surface as employed in the Cl pulse discussed earlier, but looked at the angular dependence of sputtering probability of the Ar ions as shown in Figure~\ref{fig:ar_probabilities}.
We find that the Ar impacts do not sputter away the Mo atoms at this impact energy at any incident angle. The probability of sputtering a S atom from the surface was only about 0.2\% when the impact angle is between 0 and 40 degrees. No S removal is possible at wider impact angles. We also calculated the probability of displacing S atoms from their lattice sites by Ar ion impacts and found that it is roughly between 1.2\% to 3.6\%. Thus, no continuous etching of \ce{MoS2} is possible at this ion energy at any angle of incidence.

\begin{figure}[ht]
    \centering
    \includegraphics[]{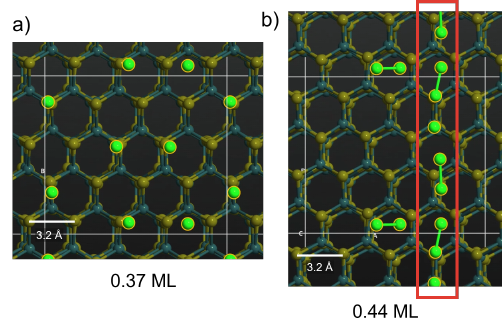}
    \caption{Panel a) shows separated Cl atoms on \ce{MoS2} surface at 0.37 ML coverage. Panel b) shows the \ce{Cl2} chain formation at a coverage of 0.44 ML. }
    \label{fig:cl-pulse}
\end{figure}

Having verified the self-limiting behavior of both reactant pulses, we attempted to simulate the sequential pulses of Cl radicals and Ar ions. For this, we used the following recipe. We first placed 1 Cl atom on the surface of \ce{MoS2} and optimized the geometry to find a local minimum and then thermalized the geometry by performing an MD at 300 K. Following that we placed another Cl at a free site on the surface and repeated the same procedure. The Cl atoms remained separated on the surface up to a coverage of 0.37 ML as shown in Figure~\ref{fig:cl-pulse} a. At 0.44 ML, the Cl atoms on the surface recombined and formed \ce{Cl2} molecules that arranged themselves into a chain formation as seen in Figure~\ref{fig:cl-pulse} b. This observation is not new as Cl is known to form such molecular chains on some  metal surfaces as well.

\begin{figure}[ht]
    \centering
    \includegraphics[]{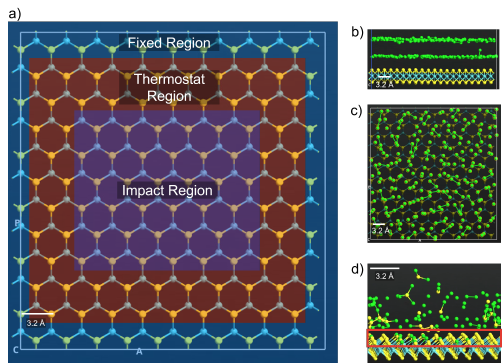}
    \caption{Panel a) shows how SPS regions are defined in a 2D \ce{MoS2} system. Panels b) and c) shows the side and top view of Cl covered \ce{MoS2} surfaces used in this study where the over layers of Cl can be seen. Panel d) shows the surface after Ar ion bombardment at 20 eV with the red rectangle highlighting the Cl atoms exchanged with top S atoms of \ce{MoS2}.}
    \label{fig:2dsps}
\end{figure}

As per the experimental evidence,\cite{kim2017atomic} Cl radicals covered up to 60 atomic percentage of the monolayer \ce{MoS2} surface after the Cl pulse and top S layer is expected to be removed first during the Ar pulse. To check this, we prepared to simulate the Ar pulse on a Cl saturated surface next. To prepare a stable Cl saturated surface of \ce{MoS2}, we needed to consider a larger surface area. Therefore, we modeled a monolayer \ce{MoS2} surface with a relatively larger surface area of 10.6 nm$^2$ and adsorbed 260 Cl atoms on it (roughly 60 percent of the number of atoms in a ML of \ce{MoS2} at this surface area) and equilibrated the system at 300 K to create a representative structure at the end of the Cl pulse as shown in Figures~\ref{fig:2dsps} b and c. For the ML \ce{MoS2}, the fixed region is set at the 4 edges of the AB plane as seen in Figure~\ref{fig:2dsps} a and the central region is used as the non-equilibrium impact region. The region between the fixed and impact regions is thermostatted at 300 K. 

Now that a representative structure at the end of the Cl pulse is generated, we proceeded to the second pulse where 1000 Ar ions at 20 eV of energy impact the surface in a sequential manner. For each impact, we simulated 400 fs of impact stage at 0.2 fs time step, 2 ps of relaxation and 2000 fbMC steps. This amounts to 32.5 ps of simulation in total per impact. The final geometry is shown in Figure~\ref{fig:2dsps} d and it can be seen that the Ar ions have split the adsorbed \ce{Cl2} species and pushed the resulting Cl atoms in to the top atomic plane of \ce{MoS2} resulting in Cl-S exchanges, a physicochemical modification of the surface layer. We can also see the formation of Cl and S containing gas by-product species in Figures~\ref{fig:2dsps} d. We can confirm from this observation that the initial phase of etching starts with replacing the surface S atoms to Cl atoms. This is consistent with the expectation\cite{kim2017atomic} that the top S atoms of \ce{MoS2} are removed first in the Ar pulse. 

\begin{figure}[ht]
    \centering
    \includegraphics[]{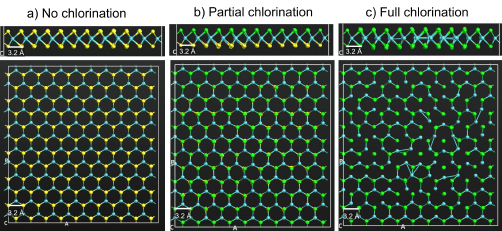}
    \caption{The images show the top and side views of three \ce{MoS2} ML systems after equilibration at 300 K: a) no chlorination, b) partial chlorination and c) full chlorination.}
    \label{fig:chlorinatedmos2}
\end{figure}

As the final investigation, we estimate the etch nucleation time depending on the level of  chlorination of the \ce{MoS2} surface. For this, we considered three cases: no chlorination, partial chlorination where the top S atoms in \ce{MoS2} layer are replaced with Cl and full chlorination where all S atoms in \ce{MoS2} layer are replaced with Cl. We prepared the representative geometries and equilibrated them at 300 K for 100 ps. The resulting geometries are shown in Figure~\ref{fig:chlorinatedmos2}. There is no significant lattice mismatch in the case of partial chlorination, but small deviations in the bond lengths are observed in the case of full chlorination which is essentially \ce{MoCl2}, a powdery material.

\begin{figure}[ht]
    \centering
    \includegraphics[]{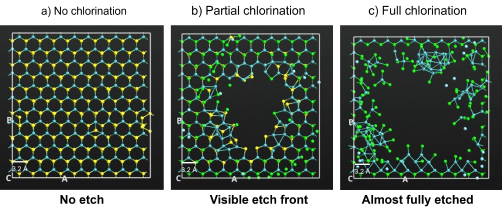}
    \caption{The images show the top view of the three Monolayer systems after 20 eV Ar ion bombardment pulse: a) no chlorination - no etch, b) partial chlorination - visible etch front, and c) full chlorination - rapid etching.}
    \label{fig:mos2etching}
\end{figure}

For the Ar pulse, the three equilibrated surfaces are exposed to 400 sequential Ar ions with a kinetic energy of 20 eV each. The ions travel in the surface normal direction from a initial distance of 10 \r{A} from the top surface and the resulting surface geometries are shown in Figure~\ref{fig:mos2etching}. 
We used the same MD settings as mentioned in the previous paragraph. It can be seen that no etching is observed in the pristine monolayer of \ce{MoS2} at this exposure (see Figure~\ref{fig:mos2etching} a). This result is consistent with the outcome for the multi-layer \ce{MoS2} surface discussed in Figure~\ref{fig:ar_probabilities}. On the contrary, an etch front is clearly visible in the case of partial chlorination (see Figure~\ref{fig:mos2etching} b) and the etch nucleation was observed on this surface already after 80 Ar impacts. For the full chlorination case, most of the material in the impact region is etched away within a few impacts as the \ce{MoCl2} units interact very weakly within the layer (see Figure~\ref{fig:mos2etching} c). We can deduce from this study that the amount of chlorine in the system after the first half cycle of this ALE process combined with the number of Cl-S exchanges in the initial phase of Ar pulse relates to the ease of etch of the modified layer in the second half cycle. Too much chlorine on the surface during the first pulse may also lead to premature chlorination of the sub-surface layers resulting in etching of multiple layers.

\section{Conclusion}
Efficiently training a machine-learned force field suitable for atomistic surface process modeling is a delicate matter. Ab-initio DFT is the standard choice for generating training data, and often times by simply sampling long ab-initio MD trajectories. However, for surface process modeling, special care must be made to sample all relevant parts of several different training data domains, and sampling MD trajectories easily becomes impractical. We have demonstrated how a different approach can efficiently build MLFFs suitable for process simulations by effectively combining structure generation protocols and active learning approaches. Also, this approach can massively reduce the amount of reference data needed to train MLFFs for complex process simulations. In this paper, we showcased two applications where custom moment tensor potentials (MTP) were trained and employed in molecular dynamics (MD) simulations to explicitly sample key reaction mechanisms taking place in \ce{HfCl4} pulse in the ALD of \ce{HfO2} and the ALE of \ce{MoS2}. We described the practical aspects of MLFF training, in particular, the batch learning and active learning approaches used in the MTP training. We presented a domain specific training data generation approach for specifically training MTPs for surface process simulation (SPS) applications. This approach allows simultaneously generating relevant training data from different domains such as gas phase, bulk and surfaces in an efficient and organized manner. Using the trained MTPs, we calculated coverage dependent adsorption probability of \ce{HfCl4} precursor molecules on a reactive \ce{HfO2} growth surface along with a theoretical estimate of maximum self-limiting molecular packing on the surface. We also simulated plasma surface interactions by modeling the high energy impacts of \ce{Cl} species on \ce{MoS2} surfaces and calculated energy dependent adsorption probability. We also estimated the threshold energy to cause physical modification of the 2D material surface by the plasma species. Synergistic effect of adsorbed Cl and Ar ions towards etching of ML \ce{MoS2} surface were identified from the detailed MD simulations. We believe that this MTP-SPS approach can support the process engineers in understanding existing processes and in discovering novel processes for challenging materials.

\section{Author Declarations}
The Authors have no conflicts to disclose.

\section{Data Availability}
The data that support the findings of this study are available from the corresponding author upon reasonable request.

\nocite{*}
\bibliography{aipsamp}

\end{document}